\documentclass[nofootinbib, reprint, amsmath,amssymb, aps]{revtex4-1}
\usepackage{graphicx}
\usepackage{bm}
\graphicspath{ {./figures/} }

\usepackage[dvipsnames]{xcolor}
\usepackage{footnote}
\usepackage{enumitem}
\usepackage{qcircuit}
\usepackage{float}
\usepackage{mathtools}
\usepackage{amsmath}
\usepackage{subcaption}
\usepackage[colorlinks]{hyperref}

\usepackage{comment}

\usepackage[bottom]{footmisc}

\usepackage{verbatim}

\usepackage{caption}

\usepackage{lineno}

\begin{document}

\title{Qubit-excitation-based adaptive variational quantum eigensolver}

\author{Yordan S. Yordanov$^{1,2} $ }
\author{V. Armaos$^3$}
\author{Crispin H. W. Barnes$^1$}
\author{David R. M. Arvidsson-Shukur$^{2, 1}$}

\affiliation{$\ ^1$ Cavendish Laboratory, Department of Physics, University of Cambridge, Cambridge CB3 0HE, United Kingdom}
\affiliation{$\ ^2$ Hitachi Cambridge Laboratory, J. J. Thomson Avenue, CB3 0HE,
Cambridge, United Kingdom}
\affiliation{$\ ^3$ Laboratory of Atmospheric Physics, Department of Physics, University of Patras, Patras, Greece}

\email{E-mail: Y.S.Yordanov - yy387@cam.ac.uk, D.R.M.A.Shukur - drma2@cam.ac.uk }

\maketitle

\onecolumngrid

\section{Abstract}

Molecular simulations with the variational quantum eigensolver (VQE) are a promising application for emerging noisy intermediate-scale quantum computers.
Constructing accurate molecular ans\"atze that are easy to optimize and implemented by shallow quantum circuits is crucial for the successful implementation of such simulations. 
Ans\"atze are, generally, constructed as
series of fermionic-excitation evolutions.
Instead, we demonstrate the usefulness of constructing ans\"atze with ``qubit-excitation evolutions'', which, contrary to fermionic excitation evolutions, obey ``qubit commutation relations''.
We show that qubit excitation evolutions, despite the lack of some of the physical features of fermionic excitation evolutions, accurately construct ans\"atze, while requiring asymptotically
fewer gates.
Utilizing qubit excitation evolutions, we introduce the qubit-excitation-based adaptive (QEB-ADAPT)-VQE protocol.
The QEB-ADAPT-VQE is a modification of the ADAPT-VQE that performs molecular simulations using a problem-tailored ansatz, grown iteratively by appending evolutions of qubit excitation operators.
By performing classical numerical simulations for small molecules, we benchmark the QEB-ADAPT-VQE, and compare it against the original fermionic-ADAPT-VQE and the qubit-ADAPT-VQE.
In terms of circuit efficiency and convergence speed, we demonstrate that the QEB-ADAPT-VQE outperforms the qubit-ADAPT-VQE, which to our knowledge was the previous most circuit-efficient scalable VQE protocol for molecular simulations.

\section{ Introduction}\label{Intro}

Quantum computers are anticipated to enable simulations of quantum systems more efficiently and accurately than classical computers \cite{QC_gen_1, QC_gen_2}.
A promising algorithm to perform this task on emerging noisy intermediate-scale quantum (NISQ) \cite{NISQ,google_supreme, vqe_nisq} computers is the variational quantum eigensolver (VQE) \cite{vqe_general, vqe_general_2, vqe_gen_3, vqe_2,VQE_accelerated, vqe_google_hf, vqas_will_fail}. 
The VQE is a hybrid quantum-classical algorithm that estimates the lowest eigenvalue of a Hamiltonian $H$ by minimizing  the energy expectation value $E(\pmb{\theta})= \langle \psi(\pmb{\theta})|H| \psi(\pmb{\theta}) \rangle$ with respect to a parametrized state $|\psi(\pmb{\theta})\rangle = U(\pmb{\theta})|\psi_0\rangle$. 
Here, $\pmb{\theta}$ is a set of variational parameters, and the unitary $U(\pmb{\theta})$ is an ansatz. 
Compared to other purely quantum algorithms for eigenvalue estimation, like the quantum-phase-estimation algorithm \cite{QC_QI,QPE}, the VQE requires shallower quantum circuits. This makes the VQE more noise resistant, at the expense of requiring a higher number of quantum measurements and additional classical post-processing.

The VQE can solve the electronic structure problem \cite{vqe_general, CCSD} by estimating the lowest eigenvalue of an electronic Hamiltonian. 
A major challenge for the practical realization of a molecular VQE simulation on NISQ computers is to construct a variationally flexible ansatz $U(\pmb{\theta})$ that: (1) accurately approximates the ground state of $H$; (2) is easy to optimize; and (3) can be implemented by a shallow circuit.

These desired qualities are  satisfied, to various levels, by several types of ans\"{a}tze. The unitary coupled cluster (UCC) type, was the first to be used for molecular VQE simulations \cite{vqe_first_uccsd}. The UCC is motivated by the classical coupled cluster theory \cite{CCSD}, and corresponds to a series of unitary evolutions of fermionic excitation operators, which we refer to as ``fermionic excitation evolutions''
(see Sec. ``Ansatz elements'').
A prominent example of a UCC ansatz is the UCC Singles and Doubles (UCCSD) \cite{UCCSD,UCCSD_2, UCCSD_3, UCCSD_4, UCCSD_bogoliubov, optimized_uccsd}, which corresponds to a series of single and double fermionic-excitation evolutions. The UCCSD has been used successfully to implement the VQE for small molecules \cite{UCCSD, vqe_first_uccsd, Eff_d_q_exc}.
Due to their physically-motivated fermionic structure, UCC ans\"{a}tze
respect the symmetries  of
electronic wavefunctions, which makes these ans\"{a}tze accurate and easy to optimize.
Even a relatively simple UCC ansatz, like the UCCSD, is highly accurate for weakly correlated systems, such as molecules near their equilibrium configuration \cite{UCCSD, vqe_first_uccsd, k-upCCGSD,trotter_error}. 
However, UCC ans\"{a}tze are general-purpose built and do not take into account details of the system of interest. They contain redundant excitation terms, resulting in unnecessarily high numbers of variational parameters as well as long ansatz circuits. 
Moreover, to simulate strongly correlated systems, UCC ans\"{a}tze require higher-order excitations and/or multiple-step Trotterization \cite{k-upCCGSD}, which creates additional overhead for the quantum hardware.

Another type of ``hardware-efficient'' ans\"{a}tze \cite{vqe_hard_eff, vqe_hard_2, exchange_ansatz, part_hole, symmetry_preserve} correspond to universal unitary transformations implemented as periodic sequences of parametrized one- and two-qubit gates. 
These ans\"{a}tze are implemented by shallow circuits, and can be highly variationally flexible. 
However, as they lack physically-motivated structure, these ans\"{a}tze require a large number of variational parameters and may suffer by vanishing energy gradients along their variational parameters, making classical optimization intractable for large molecules \cite{np_hard_optimize, hard_eff_bad_2}.
In some scenarios, this is known as the barren-plateau problem \cite{hard_eff_bad_2, barren_plateau_2, barren_plateau_3, q_neural_net}.

Recently, a number of works \cite{adapt_vqe, qubit_adapt_vqe, gene_vqe, iterative_QCC, iQCC_ILC, pruning_vqe, benchmark_adapt_vqe, iter_vqe_6, schrodiner_iter_vqe} suggested new ``iterative'' VQE protocols, which instead of using general-purpose, fixed ans\"{a}tze, construct problem-tailored ans\"{a}tze on the go.
These algorithms can construct arbitrarily accurate ans\"{a}tze that are optimized in the number of variational parameters and the ansatz circuit depth, at the expense of requiring a larger number of quantum computer measurements.
The ADAPT-VQE protocols \cite{adapt_vqe, qubit_adapt_vqe} are perhaps the most prominent family of iterative VQE protocols.
The fermionic-ADAPT-VQE \cite{adapt_vqe}, which was the first iterative VQE protocol,
constructs its ansatz by iteratively appending parametrized unitary operators, which we refer to as ``ansatz elements''. The ansatz element at each iteration is sampled from a pool of spin-complement single and double fermionic excitation evolutions, based on an energy gradient hierarchy.
The fermionic-ADAPT-VQE was demonstrated to achieve chemical accuracy ($10^{-3}$ Hartree), using an ansatz with several times fewer variational parameters, and a correspondingly shallower circuit, than the UCCSD.
In the follow-up work \cite{qubit_adapt_vqe}, the qubit-ADAPT-VQE utilizes an ansatz element pool of more variationally flexible and rudimentary Pauli string exponentials.
Due to this, the qubit-ADAPT-VQE constructs even shallower ansatz circuits than the fermionic-ADAPT-VQE, thus being, to the best of our knowledge, the currently most circuit-efficient, physically-motivated VQE algorithm. 
However, the use of more rudimentary unitary operations comes at the expense of requiring additional variational parameters and iterations to construct an ansatz for a given accuracy.

In this work, we utilize unitary operations that, despite the lack of some of the physical features captured by fermionic excitation 
evolutions, achieve the accuracy of fermionic excitations evolutions as well as the hardware efficiency of Pauli string exponentials. These operations can be used to construct circuit-efficient molecular ans\"{a}tze without incurring as many additional variational parameters and iterations, as the qubit-ADAPT-VQE.
We call these unitary operations  ``qubit excitation evolutions''.
Qubit excitation evolutions \cite{UCCSD_qubit, Eff_d_q_exc,parafermions, eff_circs} are unitary evolutions of ``qubit excitation operators'', which satisfy ``qubit commutation relations''  \cite{parafermions, eff_circs}. 
Qubit excitation evolutions can be implemented by circuits that act on fixed numbers of qubits, as opposed to fermionic excitation evolutions, which act on a number of qubits that scales at least as $O(\log_2 N_{\mathrm{MO}})$ with the number of considered molecular spin-orbitals $N_{\mathrm{MO}}$. 
We show numerically, that qubit excitation evolutions can approximate an electronic wavefunction almost as accurately as fermionic excitation evolutions can.
On the other hand, qubit-excitation evolutions enjoy higher complexity than Pauli string exponentials, thus allowing for a more rapid construction of the ansatz.
We utilize qubit excitation evolutions to introduce the qubit-excitation based adaptive variational quantum eigensolver (QEB-ADAPT-VQE) protocol. 
As the name suggests, the QEB-ADAPT-VQE is an ADAPT-VQE protocol for molecular simulations that grows a problem-tailored ansatz from an ansatz-element pool of qubit excitation evolutions.
The QEB-ADAPT-VQE also features a modified ansatz-growing strategy, which allows for a more efficient ansatz construction at the expense of a constant-factor increase of quantum computer measurement.
We benchmark the performance of the QEB-ADAPT-VQE with classical numerical simulations for small molecules: $\text{LiH}$, H$_6$ and $\text{BeH}_2$.
In Sec. ``Energy dissociation curves'', we compare the QEB-ADAPT-VQE to the standard UCCSD-VQE by presenting energy dissociation curves obtained with each of the two methods.
In Sec. ``Energy convergence'', we compare the QEB-ADAPT-VQE to the fermionic-ADAPT-VQE and to the qubit-ADAPT-VQE by presenting energy convergence plots, obtained with each of the three ADAPT-VQE protocols.

\section{Results}

\subsection{Theoretical background and  notation}\label{sec:theory}

We begin with a theoretical introduction (required for the self-completeness of the paper) and by defining our notation.
Finding the ground-state electron wavefunction $|E_0\rangle$ and corresponding energy $E_0$ of a molecule (or an atom) is known as the ``electronic structure problem" \cite{CCSD}.
This problem can be solved by solving the time-independent Schr\"odinger equation $H |\Phi_0\rangle = E_0 |\Phi_0\rangle$, where $H$ is the electronic Hamiltonian of the molecule.
Within the Born-Oppenheimer approximation, where the nuclei of the molecule are treated as motionless, $H$ can be second quantized as
\begin{equation}\label{eq:2nd_q_H}
{H} = \sum_{i,k}^{N_{_{\mathrm{MO}}}} h_{i,k}^{\ } {a}_i^\dagger {a}_k^{\ } + \sum_{i,j,k,l}^{N_{_{\mathrm{MO}}}} h_{i,j,k,l}^{\ } {a}_i^\dagger {a}_j^\dagger  {a}_k^{\ } {a}_l^{\ }.
\end{equation}
As already mentioned, $N_{_{\mathrm{MO}}}$ is the number of molecular spin-orbitals, ${a}^\dagger_i$ and ${a}_i$ are fermionic creation and annihilation operators, corresponding to the $i^{\mathrm{th}}$ molecular spin-orbital, and the factors $h_{ij}$ and $h_{ijkl}$ are one- and two-electron integrals, written in a spin-orbital basis \cite{CCSD}.
The Hamiltonian expression in equation \eqref{eq:2nd_q_H} can be mapped to quantum-gate operators using an encoding method, e.g. the Jordan-Wigner \cite{JW_encode} or the Bravyi-Kitaev \cite{BK_encode} methods. 
Throughout this work, we assume the more straightforward Jordan-Wigner encoding, where the occupancy of the $i^{{\mathrm{th}}}$ molecular spin-orbital is represented by the state of the $i^{{\mathrm{th}}}$ qubit.

The fermionic operators ${a}^\dagger_i$ and ${a}_i$ satisfy anti-commutation relations
\begin{equation}\label{eq:anti_comm}
\{a_i,a^\dagger_j\} = \delta_{i,j}, \ \ \{a_i, a_j\} = \{a_i^\dagger, a^\dagger_j\} = 0.
\end{equation}
Within the Jordan-Wigner encoding, ${a}^\dagger_i$ and ${a}_i$ can be written in terms of quantum gate operators as
\begin{equation}\label{eq:a*}
a_i^\dagger = Q_i^\dagger \prod_{r=0}^{i-1} Z_r = \frac{1}{2}(X_i-\mathrm{i}Y_i)\prod_{r=0}^{i-1} Z_r \; \; \mathrm{and} 
\end{equation}
\begin{equation}\label{eq:a}
a_i =  Q_i \prod_{r=0}^{i-1} Z_r=\frac{1}{2}(X_i+\mathrm{i}Y_i)\prod_{r=0}^{i-1} Z_r,
\end{equation}
where 
\begin{equation}\label{eq:Q*}
Q_i^\dagger \equiv \frac{1}{2}(X_i - \mathrm{i} Y_i)  \; \; \mathrm{and}
\end{equation}
\begin{equation}\label{eq:Q}
 \ Q_i \equiv \frac{1}{2}(X_i + \mathrm{i} Y_i) .
\end{equation}
We refer to ${Q_i^\dagger}$ and ${Q_i}$ as qubit creation and annihilation operators, respectively. They act to change the occupancy of spin-orbital $i$. The Pauli-$z$ strings, in equations \eqref{eq:a*} and \eqref{eq:a}, compute the parity of the state and act as exchange phase factors that account for the fermionic anticommutation of $a^\dagger$ and $a$.
Substituting equations \eqref{eq:a*} and \eqref{eq:a} into equation \eqref{eq:2nd_q_H}, $H$ can be written as 
\begin{equation}\label{eq:p_H}
{H} =  \sum_r h_r \prod_{s=0}^{N_{MO}-1} \sigma_s^r,
\end{equation}
where $\sigma_s$ is a Pauli operator ($X_s$, $Y_s$, $Z_s$ or $I_s$) acting on qubit $s$, and $h_r$ (not to be confused with $h_{ik}$ and $h_{ijkl}$) is a real scalar coefficient{\color{blue}}. 
The number of terms in equation \eqref{eq:p_H} scales as $O(N_{_{\mathrm{MO}}}^4)$.

Once $H$ is mapped to a Pauli operator representation, the VQE can be used to minimize the expectation value $E(\pmb{\theta})=\langle \psi(\pmb{\theta})|H|\psi(\pmb{\theta})\rangle$.
The VQE relies upon the Rayleigh-Ritz variational principle
\begin{equation}\label{eq:R_R}
\langle \psi(\pmb{\theta})|H|\psi(\pmb{\theta})\rangle \geq E_0 ,
\end{equation}
to find an estimate for  $E_0$. 
The VQE is a hybrid quantum-classical algorithm that uses a quantum computer to prepare the trial state $|\psi (\pmb{\theta})\rangle $ and evaluate $E(\pmb{\theta})$, and a classical computer to process the measurement data and update  $\pmb{\theta}$ at each iteration.
The trial state $|\psi (\pmb{\theta})\rangle = U(\pmb{\theta}) |\psi_0\rangle$ is generated by an ansatz, $U(\pmb{\theta})$, applied to an initial reference state $|\psi_0\rangle$.

\subsection{The ADAPT-VQE protocols}

The ADAPT-VQE protocols iteratively construct problem tailored ans\"{a}tze on the go. 
At the $m^{\mathrm{th}}$ iteration one or several unitary operators, $\{U^{(m)}_r(\theta^{(m)}_r)\}$, which we refer to as ansatz elements, are appended to the left of the already existing ansatz, $U(\pmb{\theta}^{(m-1)})$:
\begin{equation}
U(\pmb{\theta}^{(m)}) = \prod_r U^{(m)}_r(\theta^{(m)}_r) U(\pmb{\theta}^{(m-1)}) = \prod_{p=m}^1 \prod_r U^{(p)}_r(\theta^{(p)}_r).
\end{equation}
The ansatz elements, $\left\{U^{(m)}_r (\theta^{(m)}_r)\right\}$, at each iteration, are chosen from a finite ansatz element pool $\mathbb{P}$, based on an ansatz-growing strategy that aims to achieve the lowest estimate of $E(\pmb{\theta}^{(m)})$.
After a new ansatz $U(\pmb{\theta}^{(m)})$ is constructed, the new set of variational parameters $\pmb{\theta}^{(m)} = \pmb{\theta}^{(m-1)} \cup \left\{ \theta^{(m)}_r  \right\}$ is optimized by the VQE, and a new estimate for $E(\pmb{\theta}^{(m)})$ is obtained.
This iterative greedy strategy results in an ansatz that is tuned specifically to the system being simulated, and can approximate the ground eigenstate of the system with considerably fewer variational parameters and a shallower ansatz circuit, than general-purpose fixed ans\"{a}tze, like the UCCSD.

In the fermionic-ADAPT-VQE, the ansatz element pool $\mathbb{P}$ is a set of spin-complement pairs of single and double fermionic excitation evolutions. 
In the qubit-ADAPT-VQE, $\mathbb{P}$ is a set of parametrized exponentials of $XY$-Pauli strings. 
The growth strategy of the fermionic-ADAPT-VQE and the qubit-ADAPT-VQE is to add, at each iteration, the ansatz element with the largest energy gradient magnitude
\begin{equation}
\left| \frac{\partial}{\partial \theta^{(m)}} \langle \psi^{(m-1)}|U^{(m )\dagger}(\theta^{(m)}) H U^{(m)}(\theta^{(m)}) | \psi^{(m-1)} \rangle \right|_{\theta=0} , \nonumber
\end{equation}
 where $|\psi^{(m-1)}\rangle$ is the trial state at the end of the $(m-1)^{\mathrm{th}}$ iteration.
For detailed descriptions of the fermionic-ADAPT-VQE and the qubit-ADAPT-VQE, we refer the reader to Refs. \cite{adapt_vqe} and \cite{qubit_adapt_vqe}, respectively.

\subsection{Ansatz elements}\label{sec:q_and_f_exc}

Single and double fermionic excitation evolutions can construct an ansatz that approximates an electronic wavefuction to an arbitrary accuracy \cite{s_and_d_fci_1,s_and_d_fci_2}.
Single and double fermionic excitation operators, are defined, respectively, by the skew-Hermitian operators
\begin{equation}\label{eq:s_f_exc_op_aa}
T_{ik} \equiv  a^\dagger_i a_k - a^\dagger_k a_i \; \; \mathrm{and}
\end{equation}
\begin{equation}\label{eq:d_f_exc_op_aa}
T_{ijkl} \equiv  a^\dagger_i a^\dagger_j a_k a_l - a^\dagger_k a^\dagger_l a_i a_j.
\end{equation}
Single and double fermionic excitation evolutions are thus given, respectively, by the unitaries
\begin{equation}\label{eq:s_f_exc}
A_{ik}(\theta) = e^{\theta  T_{ik} } = \exp \left[ \theta(a^\dagger_i a_k - a^\dagger_k a_i)\right] \; \; \mathrm{and}
\end{equation}
\begin{equation}\label{eq:d_f_exc}
A_{ijkl}(\theta) = e^{ \theta T_{ijkl} } = \exp \left[ \theta( a^\dagger_i a^\dagger_j a_k a_l - a^\dagger_k a^\dagger_l a_i a_j)  \right].
\end{equation}
Using equations \eqref{eq:a*} and \eqref{eq:a}, for $i<j<k<l$, $A_{ik}$ and $A_{ijkl}$ can be expressed in terms of quantum gate operators as
\begin{equation}\label{eq:s_f_exc_qo}
A_{ik}(\theta)=\exp\left[\mathrm{i} \frac{\theta}{2}(X_i Y_k - Y_i X_k)\prod_{r=i+1}^{k-1}Z_r \right] \; \; \mathrm{and}
\end{equation}
\begin{align}
\label{eq:d_f_exc_qo}
A_{ijkl}(\theta) =\exp \Bigg[ \mathrm{i} \frac{\theta}{8} (X_i Y_j X_k X_l + Y_i X_j X_k X_l + Y_i Y_j Y_k X_l    +  Y_i Y_j X_k Y_l \nonumber  \\ - X_i X_j Y_k X_l  - X_i X_j X_k Y_l  - Y_i X_j Y_k Y_l - X_i Y_j Y_k Y_l ) \prod_{r=i+1}^{j-1} Z_r\prod_{r'=k+1}^{l-1} Z_{r'} \Bigg].
\end{align}
As seen from equations \eqref{eq:s_f_exc_qo} and  \eqref{eq:d_f_exc_qo}, fermionic excitation evolutions act on a number of qubits that scales as $O(N_{_\mathrm{MO}}$). Therefore, they are implemented by circuits whose size (in terms of number of $CNOT$s) also scales as $O(N_{_\mathrm{MO}})$.
We derived a $CNOT$-efficient method to construct circuits for fermionic excitations evolutions in Ref. \cite{eff_circs}.
The circuits for a single and a double fermionic excitation evolutions have $CNOT$ counts of $2(k-i) +1$ and $2(l+j-i-k)+9$, respectively.

\begin{figure}[H]
\includegraphics[width=9.5cm]{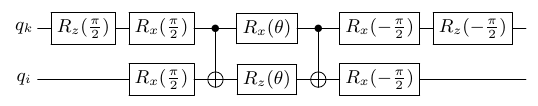}
\centering
\caption{A circuit to implement a single qubit excitation evolution. A single qubit excitation evolution is defined by the unitary operator $\tilde{A}_{ik}(\theta) = \exp \left[ \mathrm{i}\frac{\theta}{2}(X_i Y_k - Y_i X_k) \right]$, where $X$ and $Y$ are the Pauli $x$ and $y$ operators (the subscript denotes the qubit on which these operators act). $q_i$ denote the state of qubit $i$. $R_x(\theta)$ and $R_z(\theta)$ denote single-qubit rotation gates around the $x$ and $z$ axes, respectively, by $\theta$.}
\label{fig:s_q_exc}
\end{figure}

\begin{figure}[H]
\includegraphics[width=18cm]{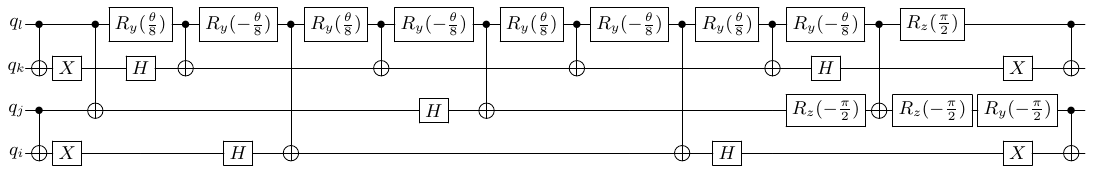}
\caption{A circuit to implement a double qubit excitation evolution. A double qubit excitation evolution is defined by the unitary operator $\tilde{A}_{ijkl}(\theta) = \exp \Bigg[ \mathrm{i} \frac{\theta}{8} (X_i Y_j X_k X_l + Y_i X_j X_k X_l  + Y_i Y_j Y_k X_l   +  Y_i Y_j X_k Y_l    - X_i X_j Y_k X_l  - X_i X_j X_k Y_l  - Y_i X_j Y_k Y_l - X_i Y_j Y_k Y_l ) \Bigg] $, where $X$ and $Y$ are the Pauli $x$ and $y$ operators. $q_i$ denote the state of qubit $i$. $H$ denotes the Hadamard gate (not to be confused with the molecular Hamiltonian), and $R_y(\theta)$ and $R_z(\theta)$ denote single-qubit rotation gates around the $y$ and $z$ axes, respectively, by $\theta$.}
\label{fig:d_q_exc_full}
\end{figure}

Qubit excitation operators are  defined by the qubit annihilation and creation operators, $Q_i$ and $Q^\dagger_i$ [equations \eqref{eq:Q*} and \eqref{eq:Q}], which satisfy the qubit commutation relations
\begin{equation}\label{eq:Q_Q}
 \{Q_i, Q_i^\dagger\} = \delta_{i,j},  [Q_i, Q^\dagger_j] = 0  \text{ if } \ i \neq j , \; \;  \mathrm{ and } \; \; [Q_i, Q_j] = [Q_i^\dagger, Q^\dagger_j] = 0 \text{ for all }i,j.  
\end{equation}
Some authors have referred to these commutation relations as parafermionic \cite{parafermions}.
Single and double qubit excitation operators are given, respectively, by the skew-Hermitian operators 
\begin{equation}\label{eq:s_q_exc_op_qq}
\tilde{T}_{ik} \equiv  Q^\dagger_i Q_k - Q^\dagger_k Q_i \; \; \mathrm{and}
\end{equation}
\begin{equation}\label{eq:d_q_exc_op_qq}
\tilde{T}_{ijkl} \equiv  Q^\dagger_i Q^\dagger_j Q_k Q_l - Q^\dagger_k Q^\dagger_l Q_i Q_j.
\end{equation}
Thus, single and double qubit-excitation evolutions are given, respectively, by the unitary operators
\begin{equation}\label{eq:s_q_exc}
\tilde{A}_{ik}(\theta) = e^{\theta  \tilde{T}_{ik} } = \exp \big[ \theta(Q^\dagger_i Q_k - Q^\dagger_k Q_i) \big] \; \; \mathrm{and}
\end{equation}
\begin{equation}\label{eq:d_q_exc}
\tilde{A}_{ijkl}(\theta) = e^{ \theta \tilde{T}_{ijkl}} = \exp \left[ \theta(Q^\dagger_i Q^\dagger_j Q_k Q_l - Q^\dagger_k Q^\dagger_l Q_i Q_j) \right].
\end{equation}
Using equations \eqref{eq:Q*} and \eqref{eq:Q}, $\tilde{A}_{ik}$ and $\tilde{A}_{ijkl}$ can be re-expressed in terms of quantum gate operators as
\begin{equation}\label{eq:s_q_exc_qo}
\tilde{A}_{ik}(\theta)= \exp \left[ \mathrm{i}\frac{\theta}{2}(X_i Y_k - Y_i X_k) \right] \; \; \mathrm{and}
\end{equation}
\begin{align}
\label{eq:d_q_exc_qo}
\tilde{A}_{ijkl}(\theta) = \exp \Bigg[ \mathrm{i} \frac{\theta}{8} (X_i Y_j X_k X_l + Y_i X_j X_k X_l  + Y_i Y_j Y_k X_l   +  Y_i Y_j X_k Y_l \nonumber \\  - X_i X_j Y_k X_l  - X_i X_j X_k Y_l  - Y_i X_j Y_k Y_l - X_i Y_j Y_k Y_l ) \Bigg] .
\end{align}
As seen from equations \eqref{eq:s_q_exc_qo} and \eqref{eq:d_q_exc_qo}, unlike fermionic excitation evolutions, qubit excitation evolutions act on a fixed number of qubits, and can be implemented by circuits that have a fixed number of $CNOT$s.
Single qubit excitation evolutions can be performed by the circuit in Fig. \ref{fig:s_q_exc}, with a $CNOT$ count of $2$. Double qubit excitation evolutions can be performed by the circuit in Fig. \ref{fig:d_q_exc_full}, which was introduced in Ref. \cite{eff_circs}, with a $CNOT$ count of $13$.

For larger systems, qubit excitation evolutions are increasingly more $CNOT$-efficient compared to fermionic excitation evolutions, whose $CNOT$ count scales as $O(N_{_\mathrm{MO}})$ in the Jordan-Wigner encoding and as $O(\log N_{_\mathrm{MO}})$ in the Bravyi-Kitaev encoding.
On the other hand, single and double qubit excitation evolutions, as seen from equations \eqref{eq:s_q_exc_qo} and \eqref{eq:d_q_exc_qo}, correspond to combinations of $2$ and $8$, mutually commuting Pauli string exponentials, respectively.  
Hence, by constructing ans\"{a}tze with qubit excitation evolutions instead of Pauli string exponentials, we  decrease the number of variational parameters.
A further advantage of qubit excitation evolutions is that they allow for the local circuit-optimizations of Ref. \cite{eff_circs}, which Pauli string exponentials do not.

When comparing the QEB-ADAPT-VQE with the fermionic-ADAPT-VQE (see Sec. ``Energy convergence''), we assume the use of the qubit- and fermionic-excitation evolutions circuits derived in Ref. \cite{eff_circs}.
To our knowledge, these are the most $CNOT$-efficient circuits for these two  types of unitary operations. 
For the qubit-ADAPT-VQE, we assume that an exponential of a Pauli string of length $l$ is implemented by a standard $CNOT$ staircase construction \cite{vqe_circs, vqe_general, eff_circs}, with a $CNOT$ count of $2(l-1)$. 
Global circuit optimization is beyond the scope of this paper.

\subsection{The QEB-ADAPT-VQE protocol}\label{sec:algorithm}

In the previous section, we formally introduced qubit excitation evolutions and presented the circuits that implement such unitary evolutions.
Here, we describe the three preparation components, and the fourth iterative component, of the QEB-ADAPT-VQE protocol. 

First, we transform the molecular Hamiltonian $H$ to a quantum-gate-operator representation as described earlier.
This transformation is a standard step in every VQE algorithm. It involves the calculation of the one- and two-electron integrals $h_{ik}$ and $h_{ijkl}$ [equation \eqref{eq:2nd_q_H}], which can be done efficiently (in time polynomial in $N_{_\mathrm{MO}}$) on a classical computer \cite{vqe_general}.

Second, we define an ansatz element pool $\mathbb{P}(\tilde{A}, N_{_\mathrm{MO}})$ of all unique single and double qubit excitation evolutions, $\tilde{A}_{ik}(\theta)$ and $\tilde{A}_{ijkl}(\theta)$, respectively, for $i,j,k,l \in \{0,N_{_\mathrm{MO}}-1\}$.
The size of this pool is $|| \mathbb{P}(\tilde{A}, N_{_\mathrm{MO}}) ||=\binom{N_{_\mathrm{MO}}}{2} + 3 \binom{N_{_\mathrm{MO}}}{4}$. Here, $|| \cdot ||$ denotes a set's cardinality.

Third, we choose an initial reference state $|\psi_0\rangle$. For faster convergence, $|\psi_0\rangle$ should have a significant overlap with the unknown ground state, $|E_0\rangle$. In the classical numerical simulations presented in this paper we use the conventional choice of the Hartree-Fock state \cite{classical_methods}.

Fourth, we initialize the iteration number to $m=1$, and the ansatz to the identity $U \rightarrow U^{(0)} = I$. Then, we initiate the QEB-ADAPT-VQE iterative loop. We start by  describing the six steps of the $m^\mathrm{th}$ iteration of the QEB-ADAPT-VQE. We then comment on these steps.

\begin{enumerate}

\item Prepare state $|\psi^{(m-1)}\rangle = U(\pmb{\theta}^{(m-1)}) |\psi_0\rangle$, with  $\pmb{\theta}^{(m-1)}$ as determined in the previous iteration.

\item For each qubit excitation evolution $\tilde{A_p}(\theta_p) =e^{\theta_p \tilde{T}_p}  \in \mathbb{P}(\tilde{A}, N_{_\mathrm{MO}})$, calculate the energy gradient:
\begin{equation}\label{eq:grad_f_exc}
\frac{\partial}{\partial \theta_p} E^{(m-1)}(\theta_p)\Big\vert_{\theta_p=0}  = 
\frac{\partial}{\partial \theta_p} \langle \psi^{(m-1)}|\tilde{A_p}^\dagger(\theta_p) H \tilde{A_p}(\theta_p) | \psi^{(m-1)} \rangle \Big\vert_{\theta_p=0}=  \langle
\psi^{(m-1)}|[ H, \tilde{T_p}] | \psi^{(m-1)} \rangle .
\end{equation}

\item Identify the set of $n$ qubit excitation evolutions, $\mathbb{\tilde{A}}^{(m)}(n)$, with largest energy gradient magnitudes. 
For $\tilde{A_p}(\theta_p) \in  \mathbb{\tilde{A}}^{(m)}(n)$:
\begin{enumerate}
\item Run the VQE to find $\min\limits_{\pmb{\theta}^{(m-1)},\theta_p} E(\pmb{\theta}^{(m-1)},\theta_p)=
 \min\limits_{\pmb{\theta}^{(m-1)},\theta_p} \langle \psi_0| U^\dagger(\pmb{\theta}^{(m-1)}) \tilde{A_p}^\dagger(\theta_p) H \tilde{A_p}(\theta_p)
U(\pmb{\theta}^{(m-1)}) | \psi_0 \rangle. $

\item Calculate the energy reduction $\Delta {E}^{(m)}_p = E^{(m-1)} - \min\limits_{\pmb{\theta}^{(m-1)},\theta_p} E(\pmb{\theta}^{(m-1)},\theta_p) $ for each $p$.

\item Save the (re)optimized values of $\pmb{\theta}^{(m-1)} \cup  \{ \theta_p \} $ as $\pmb{\theta}^{(m)}_p$  for each $p$.

\end{enumerate}

\item Identify the largest energy reduction $\Delta {E}^{(m)} \equiv \Delta {E}_{p'}^{(m)} = \max\limits_{p} \big\{ \Delta {E}_{p}^{(m)} \big\} $, and the corresponding qubit excitation evolution $\tilde{A}^{(m)}(\theta^{(m)}) \equiv \tilde{A}_{p'}(\theta_{p'})$.

If $\Delta E^{(m)} < \epsilon $, where $\epsilon>0$ is an energy threshold:
\begin{enumerate}
\item Exit
\end{enumerate}
Else:
\begin{enumerate}
\item  Append $\tilde{A}^{(m)}(\theta^{(m)})$ to the ansatz: $ U(\pmb{\theta}^{(m)}) = \tilde{A}^{(m)}(\theta^{(m)}) U(\pmb{\theta}^{(m-1)}) $

\item Set $E^{(m)} = E^{(m-1)} - \Delta {E}^{(m)}_{p'}$

\item Set the values of the new set of variational parameters, $\pmb{\theta}^{(m)} = \pmb{\theta}^{(m-1)} \cup  \{ \theta_{p'} \}$, to $\pmb{\theta}^{(m)}_{p'}$

\end{enumerate}

\item (Optional) If  the ground state of the system of interest is known, \textit{a priori}, to have the same spin as $|\psi_0\rangle$, append to the ansatz the spin-complementary of $\tilde{A}^{(m)}(\theta^{(m)})$, $\tilde{A}'^{(m)}(\theta'^{(m)})$, unless $\tilde{A}^{(m)}(\theta^{(m)}) \equiv \tilde{A}'^{(m)}(\theta'^{(m)}$:
\begin{equation}
U(\pmb{\theta}^{(m)}) = \tilde{A}'^{(m)}(\theta'^{(m)})\tilde{A}^{(m)}(\theta^{(m)}) U(\pmb{\theta}^{(m-1)}) .
\end{equation}

\item Enter the $m+1$ iteration by returning to step $1$

\end{enumerate}

We now provide some more information about the steps of the protocol.
The QEB-ADAPT-VQE loop starts by preparing the trial state $|\psi^{(m-1)}\rangle$ obtained in the $(m-1)^\mathrm{th}$ iteration.
To identify a suitable qubit excitation evolution to append to the ansatz, first we calculate (step 2) the gradient of the energy expectation value, with respect to the variational parameter of each qubit excitation evolution in $\mathbb{P}(\tilde{A}, N_{_{\mathrm{MO}}})$.The gradients are evaluated at $\theta_p=0$ because of the presumption that $|\psi_0\rangle$ is close to the ground state, which suggests that the optimized value of $\theta_p$ is close to $0$.
The gradients [equation \eqref{eq:grad_f_exc}] are calculated by measuring, on a quantum computer, the expectation value of the commutator of $H$ and the corresponding qubit excitation operator $\tilde{T}_p$, with respect to $|\psi^{(m-1)}\rangle$. 
The expression for the gradient in equation \eqref{eq:grad_f_exc} is derived explicitly in Supplementary Note 1.
Note that, Steps 1 and 2 are identical to those of the original fermionic-ADAPT-VQE.

The gradients calculated in step 2, indicate how much each qubit excitation can decrease $E^{(m-1)}$. 
However, the largest gradient does not necessarily correspond to the largest energy reduction, optimized over all variational parameters. 
In step 3, we identify the set of $n$ qubit excitation evolutions with the largest energy gradient magnitudes: $\mathbb{\tilde{A}}^{(m)}(n) \in \mathbb{P}(\tilde{A}, N_{_{\mathrm{MO}}})$. We assume that $\mathbb{\tilde{A}}^{(m)}(n)$ likely contains the qubit excitation evolution that reduces $E^{(m-1)}$ the most. For each of the $n$ qubit excitation evolutions in $\mathbb{\tilde{A}}^{(m)}(n)$, we run the VQE with the ansatz from the previous iteration to calculate how much it  contributes to the energy reduction. 
Step 3 is not present in the original fermionic-ADAPT-VQE, which directly grows its ansatz by the ansatz element with largest energy-gradient magnitude (equivalent to $n=1$).
Performing step 3 for $n>1$ further reduces the ansatz circuit at the expense of more quantum computer measurements.
A study of the performance of the QEB-ADAPT-VQE for different values of $n$ is presented in Supplementary Note 5.
The study shows that for the three molecules considered in this paper, LiH, H$_6$ and BeH$_2$, a $CNOT$ reduction between $15\%$ to $25\%$ is achieved for $n=10$.

In step 4, we pick the qubit excitation, $\tilde{A}^{(m)}(\theta^{(m)})$, with the largest contribution to the energy reduction, $\Delta {E}^{(m)}$. If $\Delta {E}^{(m)}$ is below some threshold $\epsilon > 0$, we exit the iterative loop.
If instead the $|\Delta {E}^{(m)}|> \epsilon$, we add $\tilde{A}^{(m)}(\theta^{(m)})$ to the ansatz.

If it is known, \textit{a priori}, that the ground state of the simulated system has spin zero, as the Hartree-Fock state does, we assume that qubit-excitation evolutions come in spin-complement pairs. Hence, we  append the spin-complement of $\tilde{A}^{(m)}(\theta^{(m)})$, $\tilde{A}'^{(m)}(\theta'^{(m)})$ (step 5) to the ansatz.  However, unlike the fermionic-ADAPT-VQE, the QEB-ADAPT-VQE assigns independent variational parameters to the two spin-complement excitation evolutions.
The reason for this is that qubit excitation evolutions do not account for the parity of the state. Hence, additional variational flexibility is required to obtain  the correct relative sign between the two spin-complement qubit excitation evolutions. 
Performing step 5 roughly halves the number of iterations required to construct an ansatz for a particular accuracy.

In Supplementary Note 4, we discuss the computational complexity of the QEB-ADAPT-VQE.
As a worst case estimate, the QEB-ADAPT-VQE might require as many as $O(n{N_{MO}}^{16})$ quantum computer measurements.

\subsection{Classical numerical simulations}\label{sec:results}

We perform classical numerical VQE simulations for $\text{LiH}$, H$_6$ and $\text{BeH}_2$ to compare the use of qubit and fermionic excitations in the construction of  molecular ans\"atze and to benchmark the performance of the QEB-ADAPT-VQE.
LiH and BeH$_2$ have been simulated with VQE-based protocols on real quantum computers and are often
used in the field of quantum computational chemistry to classically benchmark various VQE
protocols \cite{adapt_vqe, iQCC_ILC, UCCSD, iterative_QCC, UCCSD_2}.
Similarly to Refs. \cite{qubit_adapt_vqe, adapt_vqe}, we use H$_6$ as a prototype of a molecule with a strongly correlated ground state. 
Our numerical results are based on a custom code, designed to implement ADAPT-VQE protocols for arbitrary ansatz-element pools and ansatz-growing strategies. 
The code is optimized to analytically calculate excitation-based statevectors (see Supplementary Note 2).
The code uses the \textit{openfermion-psi4} \cite{openfermion} package to second-quantize the Hamiltonian, and subsequently to transform it to quantum-gate-operator representation.
For all simulations presented in this paper, we represent the molecular Hamiltonians in the Slater type orbital-$3$ Gaussians (STO-3G) spin-orbital basis set \cite{sto_3g, sto_3g_2}, without assuming frozen orbitals. In this basis set, $\text{LiH}$, H$_6$ and $\text{BeH}_2$, have $12$, $12$ and $14$ spin-orbitals, respectively, which are represented by $12$, $12$ and $14$ qubits.
For the optimization of variational parameters, we use the gradient-descend Broyden Fletcher Goldfarb Shannon (BFGS) minimization method \cite{bfgs} from Scipy \cite{scipy}.
We also supply to the BFGS an analytically calculated energy gradient vector (see Supplementary Note 3), for a faster optimization.
We note that in the presence of high noise levels, gradient-descend minimizers are likely to struggle to find the global energy  minimum \cite{bad_grad_descent, bad_grad_descent_2}, while direct search minimizers, like the Nelder-Mead \cite{nelder_mead}, are likely to perform better \cite{direct_search, symmetry_preserve_noise}.

\subsection{Qubit versus fermionic excitations}\label{sec:q_vs_f}

In this section, we compare qubit and fermionic excitation evolutions in their ability to construct ans\"atze to approximate electronic wavefunctions. 
Directly comparing the QEB-ADAPT-VQE and the fermionic-ADAPT-VQE (as we do in Sec. ``Energy convergence'') does not constitute a fair comparison of the two types of excitation evolutions: the QEB-ADAPT-VQE
assigns one variational parameter per qubit excitation evolution in its ansatz, whereas the fermionic-ADAPT-VQE assigns one variational parameter per spin-complement pair of fermionic excitation evolutions.
Consequently, here we compare the QEB-ADAPT-VQE for $n=1$ and step 5 not implemented, to the fermionic-ADAPT-VQE when it grows its ansatz by appending individual fermionic excitation evolutions (instead of spin-complement pairs of fermionic excitation evolutions). In this way, the two protocols differ only in using a pool of qubit excitation evolutions, and a pool of fermionic excitation evolutions, respectively.

\begin{figure}[H]
\centering
\includegraphics[width=18.5cm]{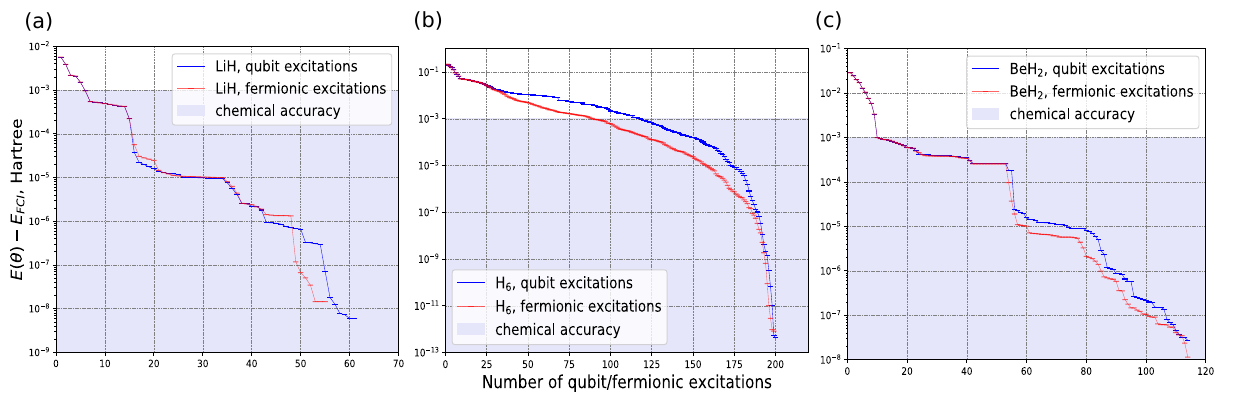}

\caption{ Comparison of qubit and fermionic excitation evolutions at equilibrium bond distances. The three subfigures present energy convergence plots for the ground states of: \textbf{(a)}  $\text{LiH}$, \textbf{(b)} H$_6$ and \textbf{(c)} $\text{BeH}_2$, in the STO-3G orbital basis set, at bond distances of $r_{Li-H}=1.546 \textup{\AA}$, $r_{H-H}=1.5 \textup{\AA}$ and $r_{Be-H}=1.316 \textup{\AA}$, respectively.
The \textbf{blue} plots are obtained with the QEB-ADAPT-VQE for $n=1$ and step 5 not implemented. The \textbf{red} plots are obtained with the fermionic-ADAPT-VQE using an ansatz element pool of non-spin-complement fermionic excitation evolutions. The plots are terminated at $\epsilon=10^{-12}$ Hartree.}
\label{fig:q_vs_f}
\end{figure}

\begin{figure}[H]
\centering
\includegraphics[width=18.5cm]{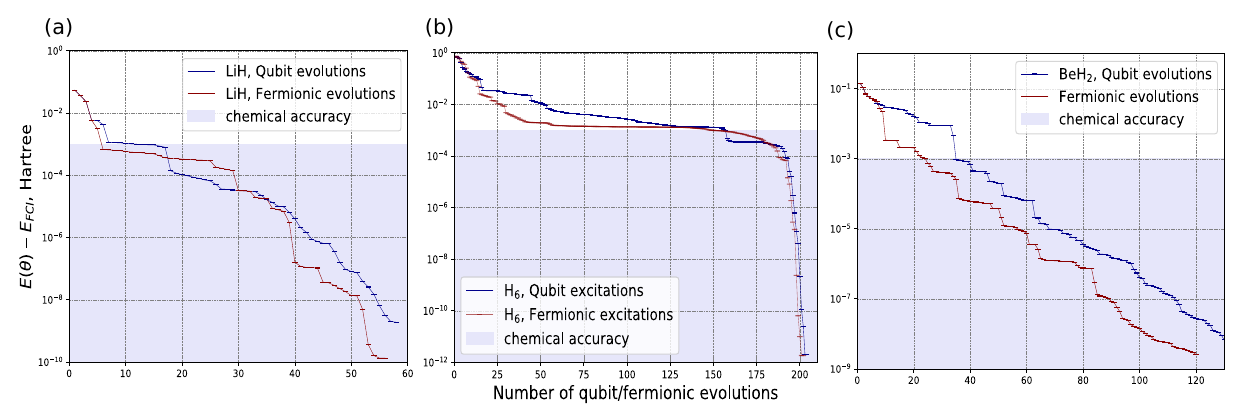}

Figure \ref{fig:q_vs_f} shows energy convergence plots, obtained with the two protocols as explained above, for the ground states of LiH (Fig. \ref{fig:q_vs_f}.a), H$_6$ (Fig. \ref{fig:q_vs_f}.b) and BeH$_2$ (Fig. \ref{fig:q_vs_f}.c) at bond distances of $r_{\text{Li-H}}=1.546 {\textup{\AA}}$, $r_{\text{H-H}}=1.5 {\textup{\AA}}$ and $r_{\text{Be-H}}=1.316 {\textup{\AA}}$, respectively. 
All plots are terminated for $\epsilon=10^{-12}$ Hartree.
The two protocols converge similarly, with the fermionic-ADAPT-VQE converging slightly faster for more than $\sim 50$ ansatz elements.
This difference is most evident for the more strongly correlated H$_6$ (Fig. \ref{fig:q_vs_f}.b), where the fermionic-ADAPT-VQE requires up to $20\%$ fewer excitation evolutions than the QEB-ADAPT-VQE to achieve a given accuracy.
These observations suggest that fermionic-excitation-based ans\"atze might be able to approximate strongly correlated states a bit better than qubit-excitation-based ans\"atze.
To further investigate this observation, in Fig. \ref{fig:q_vs_f_r3} we include energy convergence plots, similar to those in Fig. \ref{fig:q_vs_f}, but for bond distances of $r_{\text{Li-H}}=3 {\textup{\AA}}$ (Fig. \ref{fig:q_vs_f_r3}.a), $r_{\text{H-H}}=3 {\textup{\AA}}$ (Fig. \ref{fig:q_vs_f_r3}.b) and $r_{\text{Be-H}}=3 {\textup{\AA}}$ (Fig. \ref{fig:q_vs_f_r3}.c).
At larger bond distances the ground states of the LiH, and BeH$_2$ are more strongly correlated, so we expect to see larger difference in the convergence rates of the two protocols.

\caption{Comparison of qubit and fermionic excitation evolutions at large bond distances. The three subgifures present energy convergence plots for the ground states of: \textbf{(a)}  $\text{LiH}$, \textbf{(b)} H$_6$ and \textbf{(c)} $\text{BeH}_2$, in the STO-3G orbital basis set, at bond distances of $r_{Li-H}=3 \textup{\AA}$, $r_{H-H}=3 \textup{\AA}$ and $r_{Be-H}=3 \textup{\AA}$, respectively.
The \textbf{blue} plots are obtained with the QEB-ADAPT-VQE for $n=1$ and step 5 not implemented. The \textbf{red} plots are obtained with the fermionic-ADAPT-VQE using an ansatz element pool of non-spin-complement fermionic excitation evolutions. The plots are terminated at $\epsilon=10^{-12}$ Hartree.}
\label{fig:q_vs_f_r3}
\end{figure}

In Fig. \ref{fig:q_vs_f_r3}a,c we see that for LiH and BeH$_2$, at $r_{\text{Li-H}}=3 {\textup{\AA}}$ and $r_{\text{Be-H}}=3 {\textup{\AA}}$, respectively, indeed there is a larger difference in the convergence rates of the two protocols, in favour of the fermionic-ADAPT-VQE.
This is more evident for BeH$_2$ where the fermionic-ADAPT-VQE requires about $20\%$ fewer ansatz elements, on average, than the QEB-ADAPT-VQE, to achieve a given accuracy.
These results further indicate that fermionic-excitation-based ans\"atze can approximate strongly correlated states better than qubit-excitation-based ans\"atze.

\subsection{Energy dissociation curves}\label{sec:diss_curve}

\begin{figure}[H]

 \includegraphics[width=17cm]{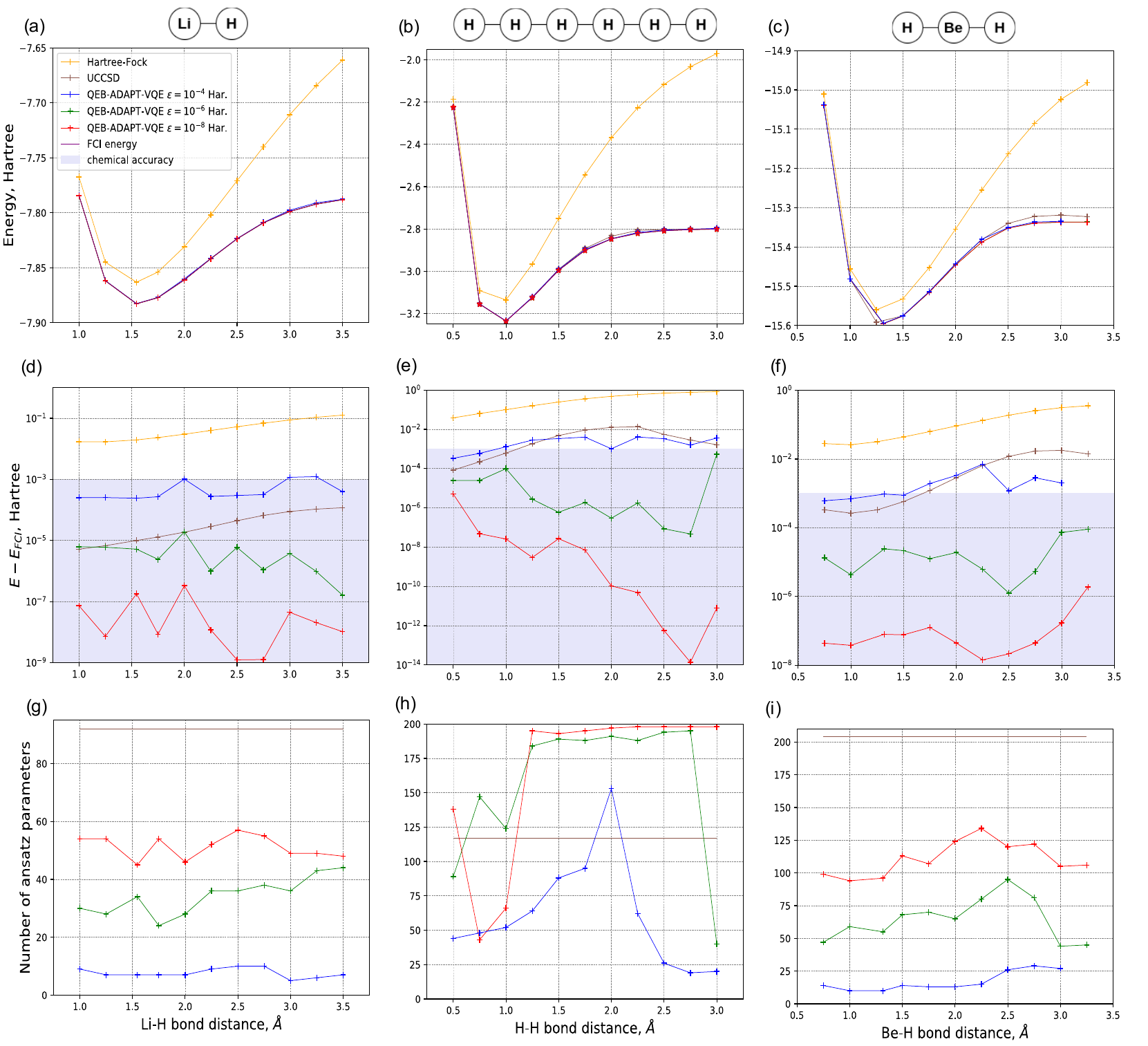}
  \caption{ Energy dissociation curves for $\text{LiH}$, H$_6$ and $\text{BeH}_2$ molecules in the STO-3G orbital basis set. \textbf{(a,b,c)} Absolute energy as function of bond distance. \textbf{(d,e,f)} Energy error with respect to the exact FCI energy as function of bond distance. \textbf{(g,h,i)} Number of ansatz variational parameters required to reach the energy accuracies in (d,e,f).
The QEB-ADAPT-VQE is performed for $n=10$ and step 5 implemented.   
   The number of variational parameters for the UCCSD is $92$, $117$ and $204$ for LiH, H$_6$ and BeH$_2$, respectively. 
  The number of variational parameters is also equivalent to the number of ansatz elements of each ansatz.
  }
\label{fig:dissociation}
\end{figure}

Figure \ref{fig:dissociation} shows energy dissociation curves for $\text{LiH}$, H$_6$ and $\text{BeH}_2$, obtained with the QEB-ADAPT-VQE for $n=10$ and energy-reduction thresholds $\epsilon_4=10^{-4}$ Hartree, $\epsilon_6=10^{-6}$ Hartree and $\epsilon_8=10^{-8}$ Hartree. Dissociation curves obtained with the Hartree-Fock (HF) method, the full configuration interaction (FCI) method, and the VQE, using an untrotterized UCCSD ansatz (UCCSD-VQE) are also included for comparison.
The UCCSD includes spin-conserving single and double fermionic evolutions only, for a fairer comparison to the QEB-ADAPT-VQE.

Figures \ref{fig:dissociation}a,b,c show the absolute values for the ground-state energy estimates. All methods except the HF, produce close energy estimates that cannot be clearly distinguished. 
In Figs. \ref{fig:dissociation}d,e,f the exact  FCI energy is subtracted in order to differentiate better the different methods and their corresponding errors.

The UCCSD-VQE achieves chemical accuracy over all bond distances for LiH (Fig. \ref{fig:dissociation}d) and over bond distances close to equilibrium configuration for H$_6$(Fig. \ref{fig:dissociation}e) and BeH$_2$(Fig. \ref{fig:dissociation}f). However, the UCCSD-VQE fails to achieve chemical accuracy for bond distances away from equilibrium configuration for H$_6$ and BeH$_2$, where the ground states become more strongly correlated.

The QEB-ADAPT-VQE for $\epsilon_4$, similarly to the UCCSD-VQE, struggles to achieve chemical accuracy for strongly correlated ground states.
However, for $\epsilon_6$ and $\epsilon_8$ the QEB-ADAPT-VQE achieves chemical accuracy over all investigated bond distances, for all three molecules. 
This indicates that the QEB-ADAPT-VQE can successfully construct ans\"{a}tze to accurately approximate strongly correlated states.

However, the real strength of the QEB-ADAPT-VQE, similarly to other ADAPT-VQE protocols, is not just in constructing accurate ans\"atze, but in constructing accurate problem-tailored ans\"atze with few variational parameters, and corresponding shallow ansatz circuits.
Figures \ref{fig:dissociation}g.h.i show plots of the number of variational parameters used by the ansatz of each method as function of bond distance. 
In the cases of LiH (Fig. \ref{fig:dissociation}g) and BeH$_2$ (Fig. \ref{fig:dissociation}i), the ans\"atze constructed by the QEB-ADAPT-VQE for $\epsilon_6$ and $\epsilon_8$ are not only more accurate than the UCCSD, but also have significantly fewer parameters. 
However, in the case of H$_6$ the QEB-ADAPT-VQE on average requires more parameters than the UCCSD. The reason for this is that H$_6$ is more strongly correlated than LiH and BeH$_2$, so even an optimally constructed ansatz would require more variational parameters than the UCCSD, to accurately approximate the ground state of H$_6$.

An interesting observation is the abrupt changes in the number of variational parameters used by the QEB-ADAPT-VQE for H$_6$ at bond distances of around $1 \text{\AA}$ , $2 \text{\AA}$, and $2.75 \text{\AA}$. The reason for these changes are molecular structure transformations, where different eigenstates of H become lowest in energy (energy-level crossings).

\subsection{Energy convergence}\label{sec:compare}

\begin{figure}[H]
\centering
\includegraphics[width=16cm]{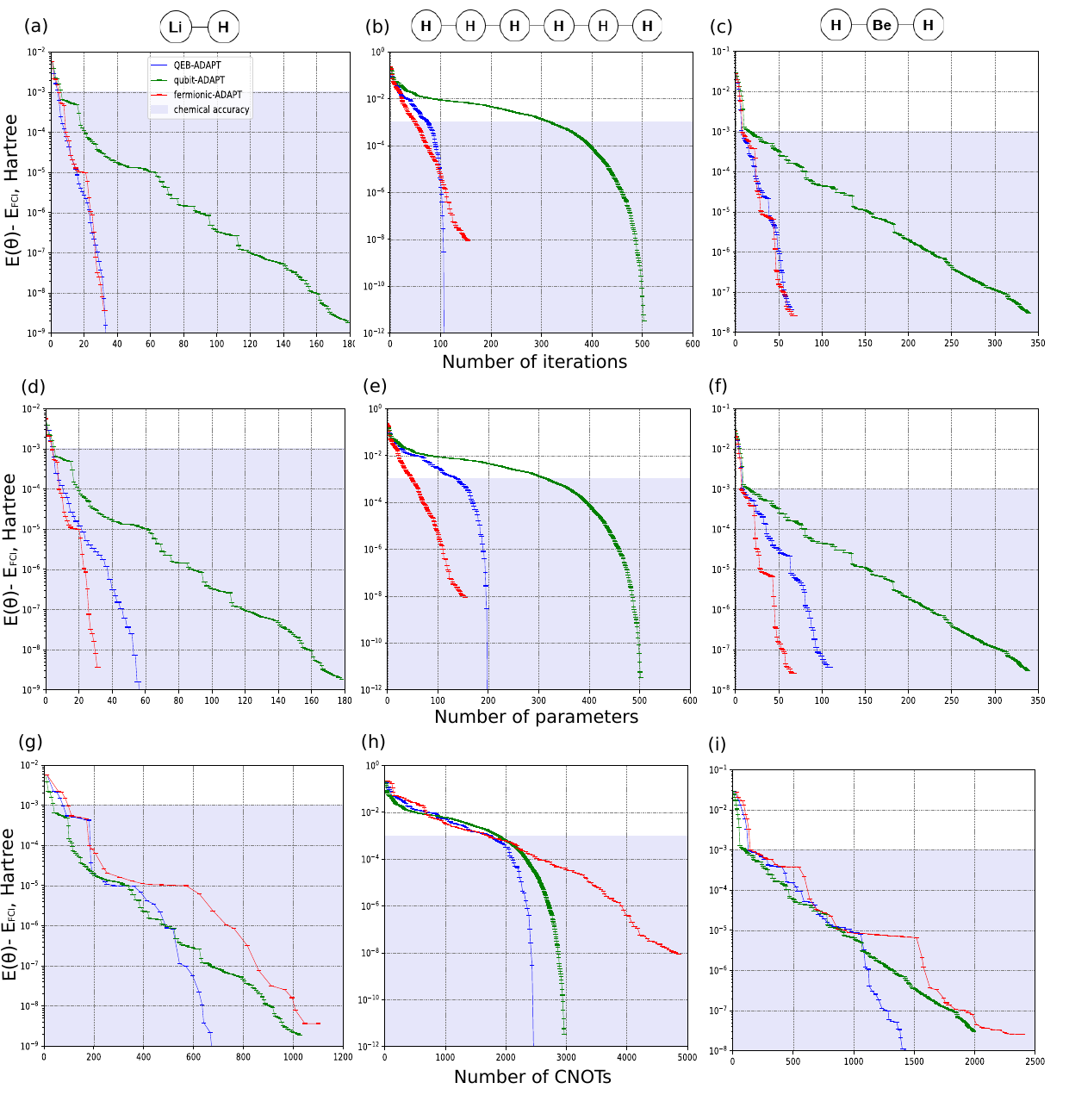}

    \caption{Comparison of the QEB-ADAPT-VQE, the fermionic-ADAPT-VQE and the qubit-ADAPT-VQE. The subfigures above present energy convergence plots for the ground states of $\text{LiH}$, H$_6$ and $\text{BeH}_2$, in the STO-3G orbital basis set, at bond distances $r_{\text{Li-H}}=1.546 {\textup{\AA}}$, $r_{\text{H-H}}=1.5 {\textup{\AA}}$ and $r_{\text{Be-H}}=1.316 {\textup{\AA}}$. 
   The plots compare the QEB-ADAPT-VQE (\textbf{blue}), the fermionic-ADAPT-VQE (\textbf{red}) and the qubit-ADAPT-VQE (\textbf{green}) protocols in terms of number of iterations \textbf{(a,b,c)}, number of parameters \textbf{(d,e,f)} and number of $CNOT$s \textbf{(g,h,i)}.   
The QEB-ADAPT-VQE is performed for $n=1$.
All convergence plots are terminated for an energy-reduction threshold of $\epsilon=10^{-12}$ Hartree.   
    The $CNOT$ counts in \textbf{g,h,i} are obtained  assuming the use of the quantum circuits discussed in Sec. ``Ansatz elements''.}
\label{fig:energy_convergence}
\end{figure}

In this section we compare the QEB-ADAPT-VQE against the fermionic-ADAPT-VQE and the qubit-ADAPT-VQE using energy convergence plots (see Fig. \ref{fig:energy_convergence}).
To ensure a fair comparison we choose the following settings for the three protocols:
We perform the QEB-ADAPT-VQE for $n=1$, using an ansatz element pool of all unique single and double qubit excitation evolutions.
The fermionic-ADAPT-VQE is performed as in Ref. \cite{adapt_vqe}, using a ansatz element pool of all unique single and double spin-complement fermionic excitation evolutions.
For the qubit-ADAPT-VQE we use an ansatz element of all evolutions of $XY$-Pauli strings of length $2$ and $4$ that have an odd number of $Y$s.
This pool consists of $O(N^4)$ Pauli string evolutions that can be combined to
obtain all qubit excitation evolutions in the ansatz element of the QEB-ADAPT-VQE (see Sec. ``Ansatz elements'').
Because of this the comparison between the QEB-ADAPT-VQE and qubit-ADAPT-VQE, in terms of
ansatz-circuit efficiency, can be considered fair.
We note that the authors of Ref. \cite{qubit_adapt_vqe} proved that the qubit-ADAPT-VQE actually can construct an ansatz that exactly recovers the FCI wavefunction, using a reduced  ansatz element pool of only $2N_{MO}-2$ Pauli string evolutions. 
This reduced pool can decrease the number of quantum computer measurements required to evaluated the energy gradients at each iteration (see step 2 of the QEB-ADAPT-VQE) from $O(N_{MO}^8)$ to $O(N_{MO}^5)$.
However, the reduced ansatz element pool will also result in a slower and less circuit-efficient ansatz construction, so using this reduced pool in the comparison with the QEB-ADAPT-VQE would not be fair.

We compare the three protocols in terms of three cost metrics, required to construct an ansatz to achieve a specific accuracy: (1) the number of iterations; (2) the number of variational parameters; and (3) the number of $CNOT$s. The number of iterations and the number of variational parameters (the number of iterations is the same as the number of variational parameters for the fermionic-ADAPT-VQE and the qubit-ADAPT-VQE, but not
for the QEB-ADAPT-VQE) determine the total number of quantum computer measurements (see Supplementary Note 4).
The $CNOT$ count of the ansatz circuit is approximately proportional to its depth. Hence, the $CNOT$ count can be used as a measure of the run time of the quantum subroutine of the VQE, which also reflects the error accumulated by the quantum hardware. 
Due to the limited coherence times of NISQ computers, the $CNOT$ count is considered as a primary cost metric.

Figure \ref{fig:energy_convergence} shows energy convergence plots, obtained with the three ADAPT-VQE protocols, for $ \text{LiH}$, H$_6$ and $\text{BeH}_2$ at bond distances of $r_{Li-H}=1.546{\textup{\AA}}$, $r_{H-H}=1.5{\textup{\AA}}$  and $r_{Be-H}=1.316{\textup{\AA}}$, respectively. 
All energy convergence plots are terminated at $\epsilon=10^{-12}$ Hartree.

In Figs. \ref{fig:energy_convergence}a,b,c we notice that the QEB-ADAPT-VQE and the fermionic-ADAPT-VQE perform similarly in terms of the number of iterations. 
This implies that the QEB-ADAPT-VQE and the fermionic-ADAPT-VQE use approximately the same number of qubit and fermionic excitation evolutions, respectively, when constructing their respective ans\"atze. 
This result is expected, because the two types of excitation evolutions perform similarly in constructing electronic wavefunction ans\"{a}tze.
Since qubit excitation evolutions are implemented by simpler circuits than fermionic excitation evolutions, the QEB-ADAPT-VQE systematically outperforms the fermionic-ADAPT-VQE in terms of $CNOT$ count in Figs. \ref{fig:energy_convergence}g,h,i.

While the QEB-ADAPT-VQE and the fermionic-ADAPT-VQE require similar numbers of iterations (Fig. \ref{fig:energy_convergence}a,b,c), the QEB-ADAPT-VQE requires up to twice as many variational parameters (Fig. \ref{fig:energy_convergence}d,e,f). 
This difference is due to the fact that the QEB-ADAPT-VQE assigns one parameter to each qubit excitation evolutions in its ansatz, whereas the fermionic-ADAPT-VQE assigns one parameter to a pair of spin-complement fermionic excitation evolutions.

Figures \ref{fig:energy_convergence}a,b,c,d show that the QEB-ADAPT-VQE converges faster, requiring systematically fewer iterations and variational parameters  than the qubit-ADAPT-VQE. 
As suggested in Sec. ``Ansatz elements'', this result is due to the fact that single and double qubit excitation evolutions correspond to combinations of $2$ and $8$ Pauli string exponentials.

In terms of CNOT count (Figs. \ref{fig:energy_convergence} g,h,i), the qubit-ADAPT-VQE is more efficient than the QEB-ADAPT-VQE at low accuracies. However, for higher accuracies, and correspondingly larger ans\"atze, the QEB-VQE-ADAPT starts to systematically outperform the qubit-ADAPT-VQE in terms of $CNOT$-efficiency. This result can be attributed to the fact that qubit evolutions allow for the local circuit optimizations introduced in Ref. \cite{eff_circs}, whereas
Pauli string evolutions, albeit more variationally flexible, do not allow for any local circuit optimizations.

As a side point, it is interesting to note that when the fermionic-ADAPT-VQE is performed with a pool of independent single and double fermionic evolutions (Figs. \ref{fig:q_vs_f} and \ref{fig:q_vs_f_r3}) it is able to converge, albeit more slowly, to higher final accuracies than when it is performed with a pool of spin-complement pairs of single and double fermionic evolutions (Fig. \ref{fig:energy_convergence}).
This is owing to the fact that the pool of independent fermionic excitation is more variationally flexible.

\section{Discussion}\label{sec:summary}

In this work, we investigated the use of qubit excitations to construct electronic VQE  ans\"atze.
We demonstrated numerically that in general an ansatz of qubit excitation evolutions can approximate a molecular electronic wavefunction almost as accurately as an ansatz of fermionic excitation evolutions.
However, fermionic-excitation-based ans\"atze were found to be slightly more accurate per number of excitation evolutions when approximating strongly correlated states.
These results suggest that, on their own, the Pauli-$z$ strings, which measure the parity of the state and account for the anticommutation of the fermionic excitation operators, play little role in the variational flexibility of an electronic wavefunction ansatz.
These results agree with previous findings in  Refs. \cite{UCCSD_qubit, qubit_adapt_vqe}.
Another advantage of fermionic excitation evolutions is that they can form spin-complement pairs of fermionic excitation evolutions.
Such spin-complement pairs can then be used to enforce parity conservation and thus reduce the number of variational parameters of an ansatz by up to a factor of $2$.
Nonetheless, fermionic excitation evolutions are implemented by circuits whose size, in terms of $CNOT$ count, scales linearly (logarithmically) in the Jordan-Wigner (Bravyi-Kitaev) encoding with the system size, as opposed to qubit excitation evolutions, which enjoy the quantum-computational benefit of being  implemented by fixed-size circuits.
Therefore, for NISQ devices, where the number of CNOTs is a primary cost factor, qubit excitation evolutions are more suitable for constructing electronic ans\"atze.

Motivated by the accuracy and circuit efficiency of qubit-excitations-based ans\"atze, we introduce the qubit-excitation based adaptive variational quantum eigensolver (QEB-ADAPT-VQE). The QEB-ADAPT-VQE simulates molecular electronic ground states with a problem-tailored ansatz, grown iteratively by appending single and double qubit excitation evolutions.
We benchmarked the performance of the QEB-ADAPT-VQE with classical numerical simulations for $\text{LiH}$, H$_6$ and $\text{BeH}_2$.
In particular, we compared the QEB-ADAPT-VQE to  the original fermionic-ADAPT-VQE, and its more slowly converging, but more circuit-efficient cousin, the qubit-ADAPT-VQE.
Compared to the fermionic-ADAPT-VQE, the QEB-ADAPT-VQE requires up to twice as many variational parameters. However, the QEB-ADAPT-VQE requires asymptotically fewer $CNOT$s, owing to its use of qubit excitation evolutions.

The simulations also showed that the qubit-ADAPT-VQE is more $CNOT$-efficient than the 
QEB-ADAPT-VQE in achieving low accuracies that correspond to small ansatz circuits. However, for higher accuracies and correspondingly larger ansatz circuits, the QEB-ADAPT-VQE systematically outperformed the qubit-ADAPT-VQE in terms of $CNOT$-efficiency.
The primary reason for this is that qubit evolutions allow for local circuit optimizations, whilst the more rudimentary Pauli string evolutions, utilized by the qubit-ADAPT-VQE, do not.
In practice, we are often just interest in reaching chemical accuracy. Therefore, one might question what is the usefulness of constructing more $CNOT$-efficient ans\"atze with the QEB-ADAPT-VQE for accuracies higher than chemical accuracy. Although the numerical results presented here are not sufficient to draw a general conclusion, they indicate that the $CNOT$-efficiency of the QEB-ADAPT-VQE becomes more evident for larger ansatz circuits. Therefore, for larger molecules, the QEB-ADAPT-VQE will likely be able to reach chemical accuracy using fewer $CNOT$s than the qubit-ADAPT-VQE.
Our simulation results also demonstrated that in terms of convergence speed, the QEB-ADAPT-VQE requires fewer variational parameters, and correspondingly fewer ansatz-constructing iterations, than the qubit-ADAPT-VQE.

These results imply that the QEB-ADAPT-VQE is more circuit-efficient and converges faster than the qubit-ADAPT-VQE, which to our knowledge was the previously most circuit-efficient, scalable VQE protocol for molecular modelling.
We do remark though, that in our comparison of the QEB-ADAPT-VQE and the qubit-ADAPT-VQE, we ignored the fact that the latter protocol can use a reduced ansatz element of $O(N_{MO})$ Pauli string evolutions, as shown in Ref. \cite{qubit_adapt_vqe}. Using a reduced ansatz element pool would decrease the number of required quantum computer measurements, but will also result in a slower and less efficient ansatz construction.
Moreover, the complexity of a single iteration of both the QEB-VQE-ADAPT and the qubit-ADAPT-VQE, might actually be dominated by running the VQE (see Supplementary Note 4).
Therefore, reducing the size of the ansatz element pool might not affect the overall complexity of the protocol.
We also note that, in theory, hardware-efficient ans\"atze and the ans\"atze of the IQCC protocol suggested in Refs. \cite{iterative_QCC, iQCC_ILC} can be implement by shallower circuits than the ans\"atze constructed by the QEB-ADAPT-VQE. However, hardware-efficient ans\"atze and the IQCC are unlikely to be scalable for large systems: the optimization of hardware-efficient ans\"atze is likely to become intractable for large systems; and the IQCC requires evaluating a number of expectation values, exponential in the number of variational parameters.

As further work, three potential upgrades to the QEB-VQE-ADAPT can be considered.
First, the ansatz element pool of the QEB-VQE-ADAPT can be expanded to include non-symmetry-preserving terms as suggested in Ref. \cite{QOCA}. Potentially, this expanded pool could further improve the speed of convergence and boost the resilience to symmetry-breaking errors of the QEB-VQE-ADAPT.
Second, methods from Ref.  \cite{pruning_vqe} can be used to ``prune'', from the already constructed ansatz, qubit excitation evolutions that have little contribution to the energy reduction. This could potentially optimize further the constructed ansatz. Third, the QEB-VQE-ADAPT functionality can be expanded to enable estimations of energies of low lying excited states. 
This will be the topic of another work (see Ref. \cite{exc_states} for a preprint).
\\

\acknowledgements
The authors wish to thank K. Naydenova and J. Drori for useful discussions.
 Y.S.Y. acknowledges financial support from the EPSRC and Hitachi via CASE studentships RG97399. D.R.M.A.-S. was supported by the EPSRC, Lars Hierta’s Memorial Foundation, and Girton College.

\section{Code availability}

The code used to perform the numerical simulations presented in this paper is publicly available at https://github.com/JordanovSJ.
Data generated during the study is available upon request from the authors (E-mail: yy387@cam.ac.uk or drma2@cam.ac.uk).

\appendix

\section{Supplementary note 1}\label{app:a_e_grad}

The expression for the single parameter energy gradient in equation (23) of the main text, can be derived as follows:

\begin{align}
\frac{\partial}{\partial \theta_p} E^{(m-1)}(\theta_p)  = 
\frac{\partial}{\partial \theta_p} \langle \psi^{(m-1)}|\tilde{A_p}^\dagger(\theta_p) H \tilde{A_p}(\theta_p) | \psi^{(m-1)} \rangle = 
\frac{\partial}{\partial \theta_p} \langle \psi^{(m-1)}| e^{\theta_p \tilde{T_p}^\dagger} H e^{\theta_p \tilde{T_p}} | \psi^{(m-1)} \rangle =  \nonumber \\
\langle \psi^{(m-1)}| e^{\theta_p \tilde{T_p}^\dagger} \tilde{T_p}^\dagger H e^{\theta_p \tilde{T_p}} | \psi^{(m-1)} \rangle +  \langle \psi^{(m-1)}| e^{\theta_p \tilde{T_p}^\dagger}  H \tilde{T_p} e^{\theta_p \tilde{T_p}} | \psi^{(m-1)} \rangle = \label{eq:grad_derive_1} \\
\langle
\psi^{(m-1)}|e^{\theta_p \tilde{T_p}^\dagger} [ H, \tilde{T_p}] e^{\theta_p \tilde{T_p}} | \psi^{(m-1)} \rangle \label{eq:grad_derive_2}
\end{align}
In going from \eqref{eq:grad_derive_1} to \eqref{eq:grad_derive_2} we use that $\tilde{T_p}$ is skew-Hermitian, so that $\tilde{T_p}^\dagger = -\tilde{T_p}$. 
In the limit $\theta_p \rightarrow 0$ expression \eqref{eq:grad_derive_2} becomes equivalent to the expression in equation (23) of the main text.

\section{Supplementary note 2}\label{app:statevector}

Here we outline the method used to calculate the trial statevectors $|\psi(\pmb{\theta})\rangle$ in the classical numerical simulations used for the results in this paper.
Calculating the trial statevectors, is the most time consuming part of the numerical simulations, and optimizing it is vital.

In this work, we are concerned with states of the form
\begin{equation}\label{eq:app_1}
|\psi(\pmb{\theta})\rangle = U(\pmb{\theta}) |\psi_0\rangle = \prod_{i=N_U}^{1} e^{\theta_i S_i} |\psi_0 \rangle,
\end{equation}
where $N_U$ is the size (the number of ansatz elements) of the ansatz $U(\pmb{\theta})$, $|\psi_0\rangle$ is the initial reference state, and $S_i$ is a skew-Hermitian operator, which in this paper corresponds to either a qubit excitation operator, a fermionic excitation operator, or a string of Pauli operators. 
Each of these three types of skew-Hermitian operators, also satisfies the relation
\begin{equation}\label{eq:s_i}
S_i^3 = -S_i.
\end{equation}

To calculate the $2^{N_{_\mathrm{MO}}}$-dimensional state-vector representing $|\psi(\pmb{\theta})\rangle$, classically, we need to calculate the $N_U$ exponents $\{e^{\theta_i S_i}\}$ in equation \eqref{eq:app_1}, and then multiply them sequentially to $|\psi_0\rangle$. Each $S_i$ is represented by an $2^{N_{_\mathrm{MO}}} \times 2^{N_{_\mathrm{MO}}}$-dimensional matrix.
Hence, the complexity of estimating each exponent $e^{\theta_i S_i}$, directly, is $O(2^{3N_{_\mathrm{MO}}})$. 
However, we can make use of relation \eqref{eq:s_i} and write each exponent in equation \eqref{eq:app_1} as
\begin{equation}\label{eq:statevector}
e^{\theta_i S_i} = \sum_{r=0}^\infty \frac{\theta_i^r S_i^r }{r!} =
I + \sum_{r_1=0}^\infty \frac{(-1)^{r_1}\theta_i^{2r_1+1} }{(2r_1+1)!}S_i + \sum_{r_2=1}^\infty \frac{(-1)^{r_2}\theta_i^{2r_2}}{(2r_2)!}S_i^2 = 
I + \sin \theta S_i + (1-\cos \theta) S_i^2.
\end{equation}
The operators $\{S_i\}$ are fixed throughout a simulation.
Therefore, if we compute in advance and store the matrix representations of each $S_i$ and $S_i^2$, we can evaluate the expression in equation \eqref{eq:statevector} by performing matrix addition only, which has a complexity of $O(2^{2 N_{_\mathrm{MO}}})$. 
Hence, the calculation of $|\psi(\pmb{\theta})\rangle$, requires $N_U$ matrix-to-vector multiplications and $N_U$ matrix additions, which gives a total complexity of $O(N_U 2^{2N_{_\mathrm{MO}}})$.

The drawback of the method outlined above is that we need to store the matrices for all $S_i$ and $S_i^2$ operators. For example, the most memory demanding simulation in this work, running the qubit-ADAPT-VQE for $\text{BeH}_2$, required around $2$GB of RAM to store the matrices for all Pauli string operators, which define the ansatz element pool of the qubit-ADAPT-VQE, and their respective squares.
However, in this case, a speed-up of nearly a factor of $20$ was achieved, in comparison to calculating $|\psi(\pmb{\theta})\rangle$ with the general IBM's Qiskit statevector simulator.

\section{Supplementary note 3}\label{app:grad_vec}

When using a gradient-descent minimizer, e.g. the BFGS, we have the option to supply a function that returns the gradient vector of the minimized function.
If we are close to the global minimum, supplying a gradient vector function guarantees a faster optimization of the variational parameters.
In the case of minimizing the Hamiltonian expectation value $E(\pmb{\theta}) =\langle \psi(\pmb{\theta})| H |\psi(\pmb{\theta})\rangle$, the $i^{\mathrm{th}}$ component of the energy gradient vector, $\pmb{\nabla} E(\pmb{\theta})$, is given by
\begin{align}
\pmb{\nabla}_i E(\pmb{\theta}) = \frac{\partial}{\partial \theta_i} \langle \psi(\pmb{\theta})| H |\psi(\pmb{\theta})\rangle = 
 \frac{\partial}{\partial \theta_i} \langle \psi_0| U^\dagger(\pmb{\theta}) H U(\pmb{\theta})|\psi_0\rangle= 
\frac{\partial}{\partial \theta_i}\langle \psi_0 | \prod^{k_1=N_U}_{1} e^{\theta_{k_1} S^\dagger_{k_1}}  H \prod_{k_2=N_U}^{1} e^{\theta_{k_2} S_{k_2}} |\psi_0 \rangle =  \\  
 \langle \psi(\pmb{\theta})| H   \prod_{k_1=N_u}^{i+1} e^{\theta_{k_1} S_{k_1}} S_i \prod_{k_1=i}^{1} e^{\theta_{k_2} S_{k_2}}| \psi_0 \rangle +  \langle \psi_0 | \prod_{k_1=1}^{i} e^{\theta_{k_1} S^\dagger_{k_1}} S_i^\dagger \prod_{k_1=i+1}^{N_U} e^{\theta_{k_2} S^\dagger_{k_2}}| H |\psi(\pmb{\theta})\rangle  = 
 2  \langle \alpha_i(\pmb{\theta})| S_i | \beta_i(\pmb{\theta}) \rangle,
\end{align}
where 
\begin{equation}
|\beta_i(\pmb{\theta}) \rangle =   \prod_{k=i}^{1} e^{\theta_{k} S_{k}}| \psi_0 \rangle \ \ \text{and}
\end{equation}
\begin{equation}
| \alpha_i(\pmb{\theta})\rangle = \prod_{k=i+1}^{N_u} e^{\theta_{k} S_{k}^\dagger} H |\psi(\pmb{\theta})\rangle,
\end{equation}
and $N_U$ is the size (the number of ansatz elements) of the ansatz $U(\pmb{\theta})$.

For the numerical simulations presented in this paper, the $N_U$ components of $\pmb{\nabla} E(\pmb{\theta})$ can be calculated with minimum number of matrix multiplications by updating $|\beta_i(\pmb{\theta}) \rangle$ and $| \alpha_i(\pmb{\theta})\rangle $ in the following way:
\begin{enumerate}

\item For $i=N_U$, initiate 
\begin{equation}
| \alpha_{N_U}(\pmb{\theta})\rangle = H |\psi(\pmb{\theta}) \rangle \ \ \text{and}
\end{equation}
\begin{equation}
|\beta_{N_U}(\pmb{\theta}) \rangle = |\psi(\pmb{\theta}) \rangle 
\end{equation}

\item For $1 < i <N_U$, update
\begin{equation}
| \alpha_{i-1}(\pmb{\theta})\rangle =  e^{\theta_{i} S_{i}^\dagger}  |\alpha_{i}(\pmb{\theta}) \rangle  \ \ \text{and}
\end{equation}
\begin{equation}\label{eq:beta}
|\beta_{i-1}(\pmb{\theta}) \rangle = (e^{\theta_{i} S_{i}})^{-1}  |\beta_{i}(\pmb{\theta}) \rangle = e^{\theta_{i} S_{i}^\dagger}  |\beta_{i}(\pmb{\theta}) \rangle,
\end{equation}
where in equation \eqref{eq:beta}, we use that $S_i$ is skew-Hermitian, so $S_i^\dagger = - S_i$.

\end{enumerate}
Assuming that we have already computed and stored the matrices of the exponentials $\{e^{\theta_i S_i}\}$, when calculating $|\psi(\pmb{\theta})\rangle$, to calculate each component of $\pmb{\nabla} E(\pmb{\theta})$ we need to perform $3$ matrix-to-vector multiplications.
Thus, overall to calculate $\pmb{\nabla} E(\pmb{\theta})$ we need to perform $3N_U$ matrix-to-vector multiplications, resulting in a total cost of $O(3 N_U 2^{2 N_{_\mathrm{MO}}})$ operations.

The cost of calculating $\pmb{\nabla} E(\pmb{\theta})$ is about $3$ times the cost of calculating $|\psi(\pmb{\theta})\rangle$. However, we find that using the energy gradient vector in the optimization subroutine of the VQE, reduces the number of VQE iterations by at least an order of magnitude, which justifies the use of the gradient vector.\\
\\

\section{Supplementary note 4}\label{sec:complexity}

Here, we consider the computational complexity in terms of number of quantum computer measurements, and total run time. The computational complexity of the QEB-ADAPT-VQE is determined by steps 2 and 3.

Given that the electronic Hamiltonian, $H$, is represented by up to $O(N_{MO}^4)$ Pauli strings (see equation (7) of the main text), calculating each gradient in step 2 would require $O(N_{MO}^4)$ quantum  computer measurements.
Since $| \mathbb{P}(\tilde{A}, N_{MO}) | \propto N_{MO}^4$, the complexity of step 2, in terms of quantum computer measurements is $O(N_{MO}^8)$.
Step 2 is completely parallelizable so if multiple quantum computers are available, its time complexity can be arbitrarily reduced down to the time required to evaluate the expectation value of a single Pauli string term, which is proportional to the ansatz circuit depth, scaling as $O(m / N_{MO})$ (a $N_{MO}$-qubits circuit of $m$ qubit excitation evolutions), where $m$ is the iteration number of the QEB-ADAPT-VQE.

Using the BFGS minimizer, optimizing ansatz $U^{(m)}\big(\vec{\theta}^{(m)}\big)$, which has $O(m)$ variational parameters, would require $O(m^2)$ VQE energy evaluations. Therefore, each VQE run in step 3 would require $O(m^2 N_{MO}^4)$ quantum computer measurement. Hence, the overall complexity of step 3 in terms of measurements would be $O(n m^2 N_{MO}^4)$.
This complexity is a worst case estimate, assuming that at each iteration, all parameters $\vec{\theta}^{(m)}$ are initialized at zero. In fact, we initiate $\vec{\theta}^{(m)}$ as $\vec{\theta}^{(m-1)} \cup {0}$, so we will need fewer VQE energy evaluations to optimize the new ansatz, $U^{(m)}\big(\vec{\theta}^{(m)}\big)$.
However, the complexity can also be higher if we use a direct search minimizer, like the Nelder-Mead, which is likely to be the case in practice, when noisy quantum hardware is used. 
Again, if multiple quantum devices are available, each of the $n$ VQE runs can be executed in parallel. Hence, the time complexity would be lower bounded by the run-time of a single VQE run, $O(m^3/N_{MO})$ (the ansatz circuit depth is $O(m/N_{MO})$ and we need to perform $O(m^2)$ VQE energy evaluations).

Overall, the QEB-ADAPT-VQE would require $O\big(N_U(N_{MO}^8 + n N_U^2N_{MO}^4)\big)$ quantum computer measurements, and its run-time complexity would be lower bounded by $O(N_U^4/N_{MO})$.  
The size of the ansatz, $N_U$, depends on the desired accuracy, and also is problem specific. Therefore, it is difficult to predict how it would scale with $N_{MO}$. 
For strongly correlated states, achieving chemical accuracy might require an ansatz that consist of as many as $O(N_{MO}^4)$ qubit excitation evolutions. 
However, for weakly correlated states, the scaling of $N_U$ with $N_{MO}$ is likely to be lower.
Assuming the worst case scenario, the time complexity of the QEB-ADAPT-VQE will be lower-bounded by $O(N_{MO}^{15})$ and it will require $O(nN_{MO}^{16})$ quantum computer measurements.
For comparison, the UCCSD-VQE has a time complexity of $O(N_{MO}^{11})$, assuming maximum parallelization, and requires $O(N_{MO}^{12})$ quantum computer measurements.

\section{Supplementary note 5}\label{sec:n=1_vs_n=10}

Here, we investigate the performance of the QEB-ADAPT-VQE for different values of $n$, the number of qubit excitation evolutions considered in step 3 of the QEB-ADAPT-VQE. 
As we increase $n$, we increase the chance to pick at each iteration the qubit excitation evolution that, added to the ansatz, achieves largest energy reduction.
Following this greedy strategy is no guarantee for an optimal ansatz, since qubit evolutions do not commute in general. Nevertheless, we do expect, on average, to construct a more circuit-efficient ansatz by increasing $n$ up to some saturation value.

To test this presumption we perform classical numerical simulations to obtain energy convergence plots for the ground states of LiH, H$_6$ and BeH$_2$ in the STO-3G basis.
The simulations for the three molecules are performed for bond distances $r_{Li-H}=3 \text{\AA}$, $r_{H-H}=3 \text{\AA}$ and $r_{Be-H}=3 \text{\AA}$, away from equilibrium configurations, where correlation effects are stronger, and the effect of increasing $n$ should be more evident.
The simulation results are presented in Supplementary figure \ref{fig:iqeb(n)}.

\begin{figure}[H]
\centering
\includegraphics[width=18cm]{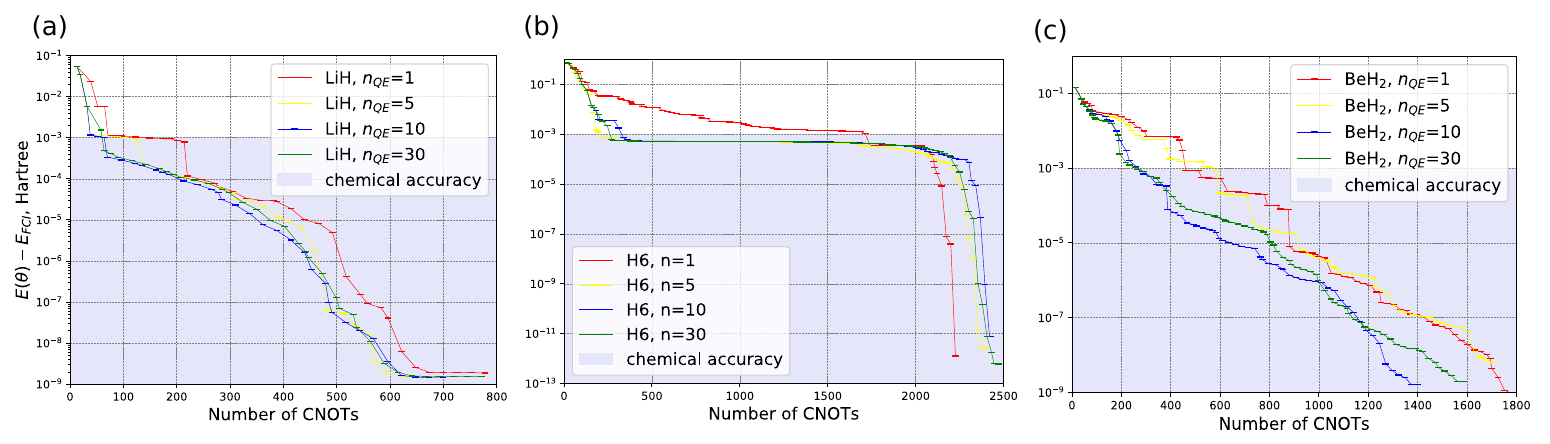}

\caption{Performance of the QEB-ADAPT-VQE for different values of the protocol parameter $n$. The subfigures above present energy convergence plots obtained with the QEB-ADAPT-VQE with different values of $n$ for the ground states of LiH (\textbf{a}), H$_6$ (\textbf{b}) and BeH$_2$ (\textbf{c}) in the SO-3G basis, at bond distances of $r_{Li-H}=3 \text{\AA}$, $r_{H-H}=3 \text{\AA}$ and $r_{Be-H}=3 \text{\AA}$. The plots are terminated at $\epsilon=10^{-12}$ Hartree.}
\label{fig:iqeb(n)}
\end{figure}

The table below summarizes the average (over number of qubit excitation evolutions) $CNOT$ count reductions, with respect to $n=1$, for each molecule and different value of $n$:
\begin{table}[H]
\centering
\begin{tabular}{c |c|c|c}
& $n=5$ & $n=10$ & $n=30$\\
\hline
LiH & $16\%$ & $20\%$ & $16\%$ \\
BeH$_2$ & $3\%$ & $26\%$ & $22\%$  \\
H$_6$ & $15\%$ & $12\%$ & $13\%$  
\end{tabular}
\caption{Average (over number of qubit excitation evolutions) $CNOT$ count reduction for QEB-ADAPT-VQE($n_{qe}>1$) as compared to QEB-ADAPT-VQE($n_{qe}=1$).}
\label{table:iqeb(n)}
\end{table}
For LiH (Supplementary figure \ref{fig:iqeb(n)}.a), the QEB-ADAPT-VQE clearly constructs ansatz circuits with fewer $CNOT$s as $n$ is increased above $1$. For BeH$_2$ (Supplementary figure \ref{fig:iqeb(n)}.c), a significant $CNOT$ count reduction is obtained for $n=10$ and $n=30$, but not for $n=5$. For H$_6$ (Supplementary figure \ref{fig:iqeb(n)}.b), the average $CNOT$ reduction is about the same for $n=5$, $n=10$ and $n=30$, but strangely the ansatz constructed by the QEB-ADAPT-VQE for $n=1$ is the most $CNOT$-efficient for accuracies higher than $10^{-4}$ Hartree.
Also, for all three molecules we observe no further $CNOT$ reduction for $n=30$ as compared to $n=10$. Actually for $n=30$ the $CNOT$ reduction is a bit lower.
 As noted above these inconsistencies can be explained by the fact that the greedy strategy to obtain the lowest estimate for $E(\vec{\theta})$ at each iteration is no guarantee for constructing an optimal ansatz, because qubit excitation evolutions do not commute.

Nonetheless, there is a clear advantage in terms of $CNOT$ count, in performing step 3 of the QEB-ADAPT-VQE for $n>1$. Despite the associated overhead in the number of quantum computer measurements with increasing $n$, this is justified as long as the bottleneck of NISQ computers is the quantum gate fidelity.
Furthermore, we can expect the $CNOT$ count reduction for $n>1$ to increase for larger molecules, because the QEB-ADAPT-VQE will have to consider a larger ansatz element pool.

\bibliographystyle{apsrev4-1}
\bibliography{references}

\end{document}